\documentclass[letter]{aa} 
\usepackage{graphicx}
\usepackage{txfonts}
\usepackage{natbib}
\bibpunct{(}{)}{;}{a}{}{,} 
\begin{document}
   \title{A fundamental plane for field star--forming galaxies}

   \author{M.A. Lara-L\'opez
          \inst{1,2}
          \and
          J. Cepa\inst{1,2}
	  \and
          A. Bongiovanni\inst{1,2}
	  \and
	  A.M. P\'erez Garc\'{\i}a\inst{1,2}
	  \and
	  A. Ederoclite\inst{1,2}
	  \and
          H. Casta\~neda\inst{1,3}
	  \and
          M. Fern\'andez Lorenzo\inst{1,2}
	  \and
          M. Povi\'c\inst{1,2}
	  \and
          M. S\'anchez-Portal\inst{4}
         }

   \institute{Instituto de Astrof\'{\i}sica de Canarias, 38200 La Laguna, Spain
              \email{mall@iac.es}
         \and
             Departamento de Astrof\'{\i}sica, Universidad de la Laguna, Spain
          \and
            Departamento de F\'{\i}sica, Escuela Superior de F\'{\i}sica y Matem\'atica, IPN, M\'exico D.F., M\'exico
          \and
             Herschel Science Center, INSA/ESAC, Madrid, Spain
   }

   \date{Received 15 April 2010 ; accepted}

\abstract  
{Star formation rate (SFR), metallicity and stellar mass are within the important parameters of star--forming (SF) galaxies that characterize their formation and evolution. They are known to be related to each other at low and high redshift  in the mass--metallicity, mass--SFR, and metallicity--SFR relations.} 
{In this work we demonstrate the existence of a plane in the 3D space defined by the axes SFR [log(SFR)(M$_{\odot}$yr$^{-1}$)], gas metallicity [12+log(O/H)], and stellar mass [log(M$_{\rm star}$/M$_{\odot}$)] of SF galaxies.}
{We used SF galaxies from the $``$main galaxy sample$"$ of the Sloan Digital Sky Survey--Data Release 7 (SDSS --DR7) in the redshift range $0.04 < z < 0.1$ and $r-$magnitudes between 14.5 and 17.77. Metallicities, SFRs, and stellar masses were taken from the Max-Planck-Institute for Astrophysics--John Hopkins University (MPA-JHU) emission line analysis database.}
{From a final sample of 32575 galaxies, we find for the first time a fundamental plane for field galaxies relating the SFR, gas metallicity, and stellar mass for SF galaxies in the local universe. One of the applications of this plane would be estimating stellar masses from SFR and metallicity. High redshift data from the literature at redshift $\sim$0.85, 2.2, and 3.5, do not show evidence for evolution in this fundamental plane.}
{}


   \maketitle
%

\section{Introduction}

Relations between important properties of astrophysical objects often lead to the discovery of the so--called fundamental planes when three parameters are involved. The fundamental plane (FP) for elliptical galaxies \citep{Djor87,Dressler87}, relates their luminosity, velocity dispersion (dynamics), and scale length (morphology). This FP represents an important tool to investigate the properties of early type and dwarf galaxies, to perform cosmological tests, and compute cosmological parameters. It is also an important diagnostic tool for galaxy evolution and mass--to--light ($M/L$) variations with redshift.

Fundamental planes have also been defined for  globular clusters  \citep{Meylan97} and galaxy clusters \citep{Schaeffer93, Adami98}. The parameter space of globular clusters, elliptical galaxies, and galaxy clusters is properly described by a geometrical plane $L \propto R^{\alpha}\sigma^{\beta} $, where $L$ is the optical luminosity of the system, $R$ is a measure of the size scale, $\sigma$ is the velocity dispersion of the system, and $\alpha$ and $\beta$ are free parameters. The FP for globular clusters, elliptical galaxies, and galaxy clusters have very similar slopes, which means that, accounting for differences in zero points, a single FP with a range of about nine orders of magnitude in luminosity can be defined \citep{Schaeffer93,Ibarra09}. 

The FP that we introduce here relates three fundamental parameters: the SFR [log(SFR)(M$_{\odot}$yr$^{-1}$)], gas metallicity [12+log(O/H)], and stellar mass [log(M$_{\rm star}$/M$_{\odot}$)] of field SF galaxies. All these variables have been related in the past by the mass--metallicity ($M-Z$) relation  \citep{Lequeux79}, the mass--SFR relation \citep{Brinchmann04}, and the metallicity--SFR relation \citep{Lara10,Lopez-Sanchez10}. Also, some authors have studied the inter-dependence of those variables \citep[e.g.][]{Hoopes07, Ellison08, Mannucci10}. In the present work, we propose the generalization of those relations defining a plane formed by a linear combination of two of those variables with respect to the third one.

The $M-Z$ relates the mass and metallicity of galaxies, with massive galaxies showing higher metallicities than less massive ones, and it has been well established for the local universe ($z \sim$ 0.1) by the work of \citet{Tremonti04} using SDSS data. The $M-Z$ relation has also been studied at low redshifts $z \sim 0.35$ \citep{Lara09a,Lara09b}, at intermediate redshifts $z \sim 0.7$ \citep[e.g.,][]{Rodrigues08}, and at high redshift $z \sim 2.2$ and $z \sim 3.5$ \citep[][respectively]{Erb06,Maiolino08}.

The stellar mass of SF galaxies is also related to the SFR, in the sense that more massive galaxies show higher SFRs \citep{Brinchmann04,Salim05}. However, \citet{Brinchmann04} emphasized that at log(M$_{\rm star}$/M$_{\odot}$) $\gtrsim$ 10, the distribution of SFRs broadens significantly and the correlation between stellar mass and SFR breaks down. At higher redshifts, \citet{Noeske07} showed the existence of a $``$main sequence$"$ (MS) for this relation over the redshift range $0.2 < z < 1.1$, with the slope of the MS  moving as a whole as $z$ increases.

The metallicity and SFR of SF galaxies are weakly correlated, as will be observed in Fig. 1. However, and despite of the high scatter, SFR increases with metallicity \citep{Lara10,Lopez-Sanchez10}.

This paper is structured as follows: Sect. 2  describes the data selection as well as the SFRs, metallicities and stellar masses estimations given by the Max-Planck-Institute for Astrophysics--John Hopkins University (MPA-JHU) group\footnote{http://www.mpa-garching.mpg.de/SDSS} and adopted in this work. In Sect. 3 we define the FP for field galaxies, and conclusions are given in Sect. 4.

\section{Data processing and sample selection}

Our study was carried out with galaxies from SDSS--DR7 \citep{York00,Abaza09}. Data were taken with a 2.5 m telescope located at Apache Point Observatory \citep{Gunn06}, further technical details can be found in Stoughton et al. (2002).

We used the emission--line analysis of SDSS--DR7 galaxy spectra performed by the MPA-JHU group. From the full dataset, we only consider objects classified as galaxies in the $``$main galaxy sample$"$ \citep{Strauss02} with apparent Petrosian $r$ magnitude in the range $14.5 < r < 17.77$  and redshift range $0.04 < z < 0.1$, that represent a complete sample in magnitude and redshift \citep{Kewley08,Kewley06}. The lower limit in redshift ensures covering $>$ 20$\%$ of the galaxy light, which is the minimum required to avoid domination of the spectrum by aperture effects \citep{Kewley05}. Following \citet{Kewley08} and \citet{Kobulnicky99}, for reliable metallicity estimates we selected galaxies with a signal--to--noise ratio (SNR) higher than 8 for the {H$\alpha$}, {H$\beta$}, [{N\,\textsc{ii}}] $\lambda$6584, [{O\,\textsc{ii}}] $\lambda$3727, [{O\,\textsc{iii}}] $\lambda$5007, and [{S\,\textsc{ii}}] $\lambda \lambda$ 6717, 6731 lines. However, using less restrictive criteria does not affect the relations derived here, but only increase their dispersion. For a detailed analysis on the line SNR see \citet{ Brinchmann04}. Finally, SF galaxies were selected following the criterion given by \citet{Kauf03a} for the BPT empirical diagnostic diagram log[{O\,\textsc{iii}}] $\lambda$5007/{H$\beta$} $\le$ 0.61/$\{$log([{N\,\textsc{ii}}] /{H$\alpha$})-0.05$\}$ + 1.3. From this final sample of 32575 galaxies, metallicities, stellar masses, and SFRs used in the present work were obtained by the MPA-JHU group following the methods described below. Since field galaxies are the dominant population of this sample, the FP presented here would be representative of field galaxies, and different from the known FP of elliptical and clusters of galaxies.

Metallicities were estimated statistically using Bayesian techniques according to \citet{Tremonti04}, based on simultaneous fits of all the most prominent emission lines ([{O\,\textsc{ii}}], {H$\beta$}, [{O\,\textsc{iii}}], {H$\alpha$}, [{N\,\textsc{ii}}], [{S\,\textsc{ii}}]) using a model designed for the interpretation of integrated galaxy spectra \citep{Charlot01}. Since the metallicities derived with this technique are discreetly sampled, they show small random offsets \citep[see for details][]{Tremonti04}. Any dependence of SFR on the estimated metallicity would be minor \citep[][]{Tremonti04,Brinchmann08}. For this work, we selected galaxies with 12+log(O/H) $>$ 8.4, corresponding to the upper branch of the R$_{23}$. However, galaxies with 12+log(O/H) $<$ 8.4, corresponding to the lower branch of the R$_{23}$ calibration, are poorly sampled, see Fig. 6 of \citet{Tremonti04} and Fig. 1.1 of \citet{Kewley08}. Therefore, to avoid a systematic dispersion in the FP, and to work with an homogeneous sample, we selected galaxies with 12+log(O/H) $>$ 8.4, which correspond to the $\sim$99$\%$ of our SF sample.

Total stellar masses were estimated from fits to the photometry using the same modelling methodology as described in Kauffmann et al (2003), with only small differences with respect to previous data released.


Finally, total SFRs for SF galaxies are derived directly from the emission lines, based on the careful modelling discussed in \citet{Brinchmann04}, who modeled the emission lines in the galaxies following the  \citet{Charlot01} prescription, obtaining a robust dust correction. Also, the metallicity dependence of the Case B {H$\alpha$}/{H$\beta$} ratio is taken into account as well. The \citet{Brinchmann04} method offers a more robust SFR estimate than using, for example, a fixed conversion factor between {H$\alpha$} luminosity and SFR \citep[e.g.][]{Kennicutt98}.

The FP presented in this study was initially identified by us using STARLIGHT data \citep{Cid05,Mateus06} for the above described sample, but estimating SFRs using the H$\alpha$ luminosity and the \citet{Kennicutt98} relation, and metallicities following the calibration of \citet{Tremonti04}. Nevertheless, although the FP derived  is the same, it is noteworthy that the plane has a lower scatter when using the robust SFR and metallicity estimations derived by the MPA-JHU group.

\section{The fundamental plane}

As mentioned in the introduction, the SFR, stellar mass, and gas metallicity of SF galaxies are related to each other. Their strong relation is evident when these data are plotted in a 3D space with ortogonal coordinate axes defined by these parameters. A careful inspection of this 3D representation, shows the existence of a plane (see Appendix A).

\begin{figure*}[t!]
 \centering
\includegraphics[scale=0.43]{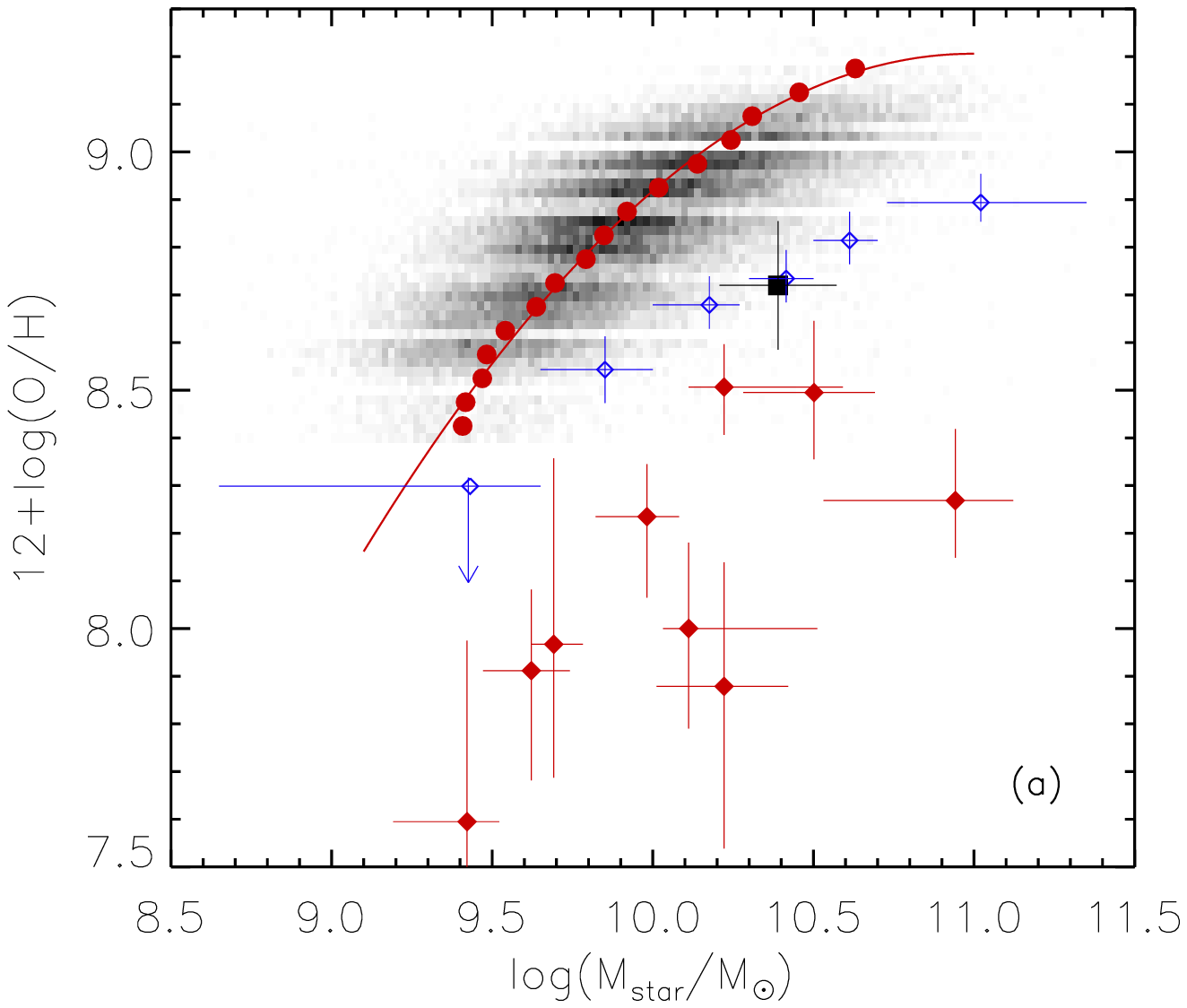}
\includegraphics[scale=0.43]{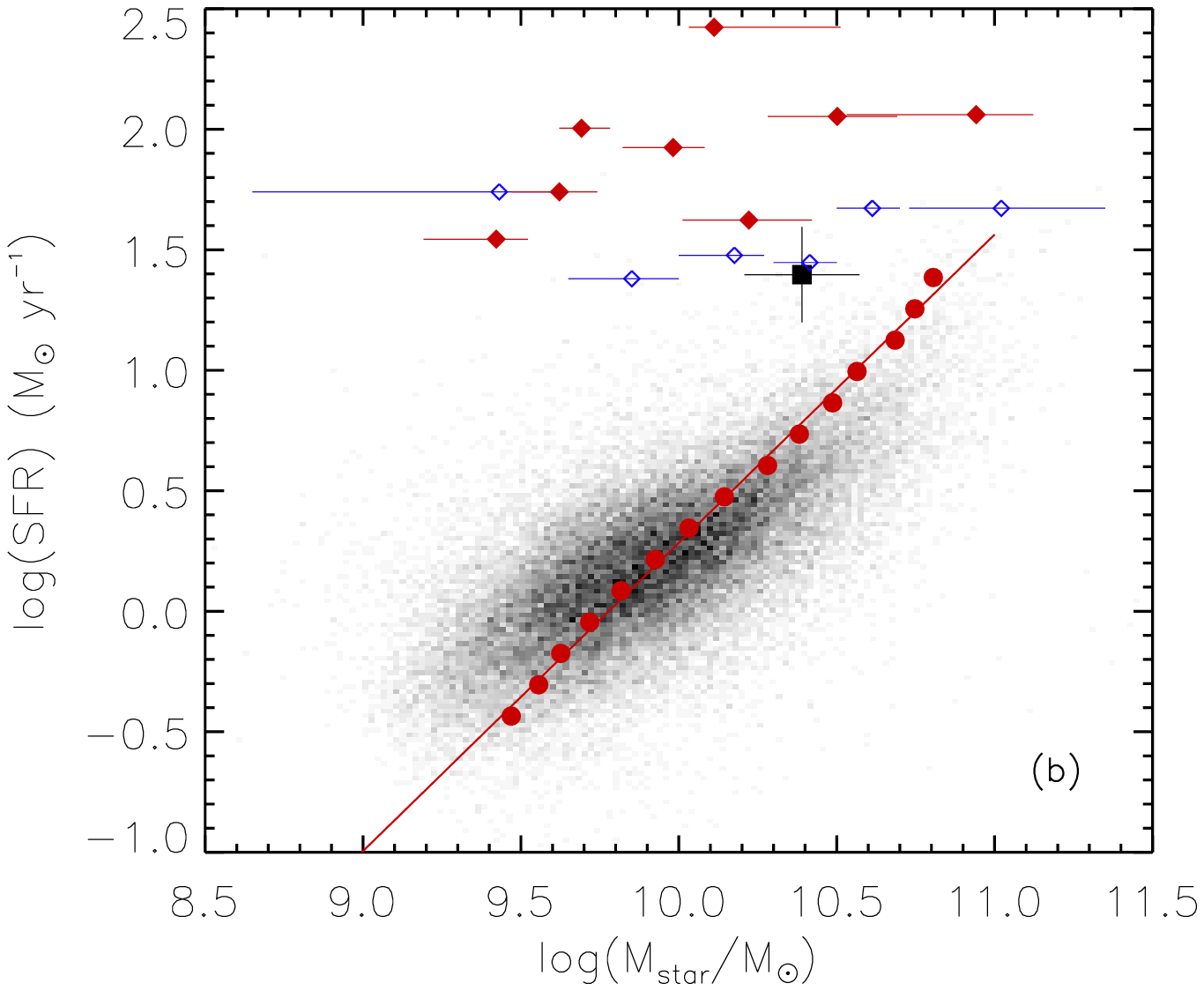}
\includegraphics[scale=0.43]{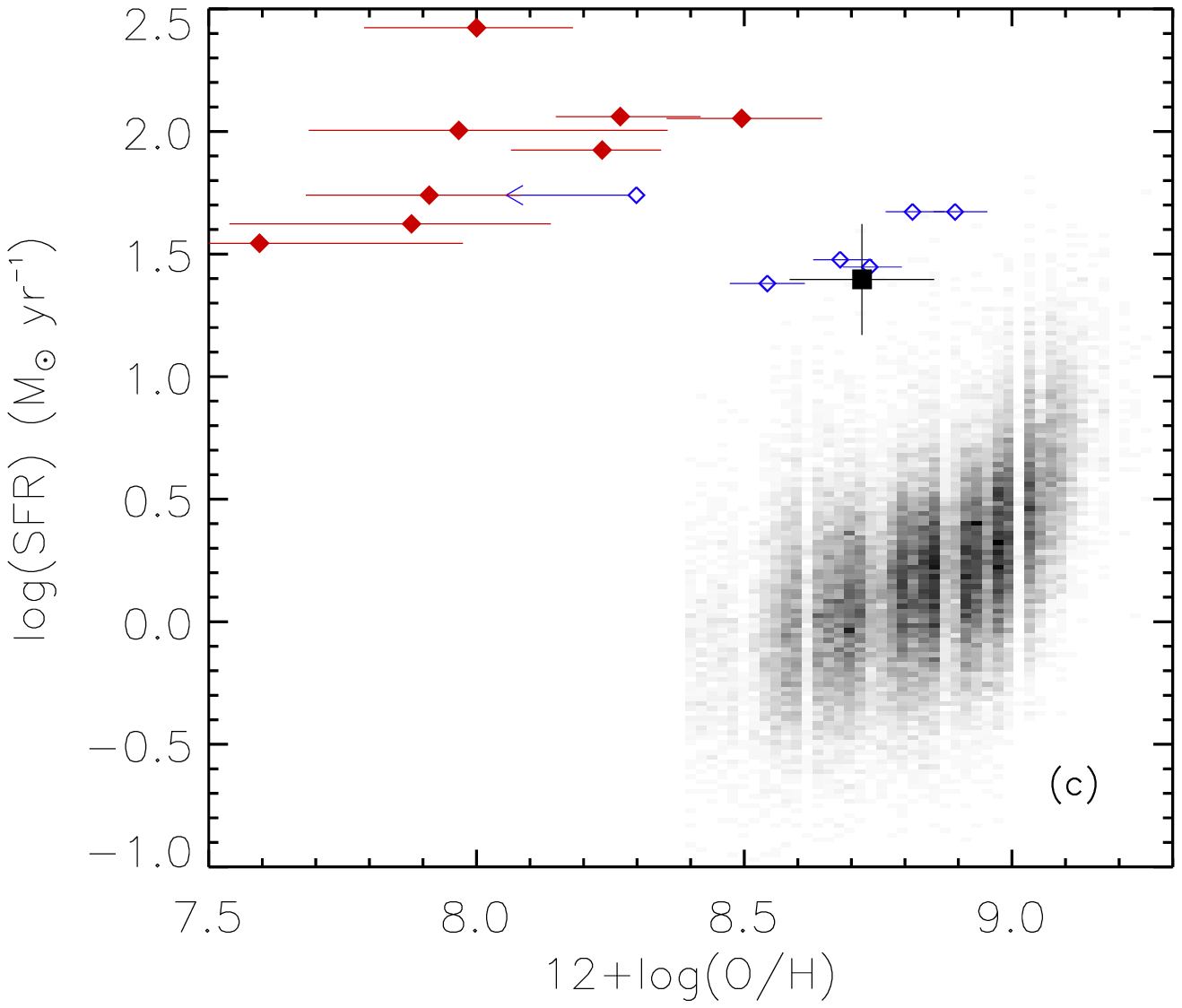}
\caption{In panels a, b, and c, the $M-Z$, stellar mass--SFR, and metallicity--SFR relations are respectively shown. The density plots are represented in bins of 0.02 dex. Red circles represent the median mass in bins of 0.05 dex in 12+log(O/H) and of 0.13 dex in log(SFR). Blue empty diamonds represent the \citet{Erb06} data at z$\sim$2.2, the bars of their data show the metallicity error and the stellar mass range of each bin, while the SFR errors are of the order of $\sim$1.3 dex. Red filled diamons represent the sample of \citet{Maiolino08} at z$\sim$3.5, the bars of their data show the metallicity and mass errors. The black filled square represents the median of the data of \citet{Rodrigues08} at z$\sim$0.85 with its respective error bars. Solid lines in panels (a) and (b) represent a two and one order polynomial fits to the median values, respectively.}
\end{figure*}

The projection of galaxy data over any pair of the axes of this 3D space will reduce to the $M-Z$, metallicity--SFR, and stellar mass--SFR relations, as shown in Fig. 1. \citet{Tremonti04} demonstrated that the metallicity of galaxies increases with the stellar mass in a relatively steep way from $10^{8.5}$ to $10^{10.5}$ M$_{\odot}$, but flattens above $10^{10.5}$ M$_{\odot}$ (see Fig. 1a). \citet{Kewley08} recalibrated the $M-Z$ relation of  \citet{Tremonti04} using the completeness criteria of redshift and magnitud given in Sect. 2. However, as the \citet{Kewley08} fit (see Fig. 1.1 of that paper) departs for the low mass population, we fit a second order polynomial to the median mass in metallicity bins of the $M-Z$ relation, 12+log(O/H)$=a_0+a_1$[log$({\rm M_{star}}/{\rm M_{\odot}})$]$+a_2$[log$({\rm M_{star}}/{\rm M_{\odot}})]^2$, with $a_0=-25.93923$, $a_1=6.39283$, $a_2=-0.29071$, and $\sigma=0.26$. Regarding the mass--SFR relation (see Fig. 1b), the SFR increases with stellar mass up to $\sim 10^{10}$ M$_{\odot}$ \citep{Brinchmann04}, while for higher mass values the scatter increases (see Fig. 1b). We fitted a line to the median mass in log(SFR) bins of the mass--SFR relation, log(SFR)$=a_0+a_1$[log$({\rm M_{star}}/{\rm M_{\odot}})$], with $a_0=-12.50704$, $a_1=1.27909$, and $\sigma=0.27$. Note that the $\sigma$ given is that of the horizontal axes [log$({\rm M_{star}}/{\rm M_{\odot}})$] in both cases.
However, for 12+log(O/H) $\lesssim 8.9$, the SFR is not strongly correlated with metallicity (see Fig 1c), whereas for higher metallicity values, the SFR increases rapidly. Therefore, since the metallicity--SFR relation does not correlate at all, this leads us towards a possible solution: a linear combination of the metallicity and SFR as a function of the stellar mass.

\begin{figure}[h!]
 \centering
\includegraphics[scale=0.43]{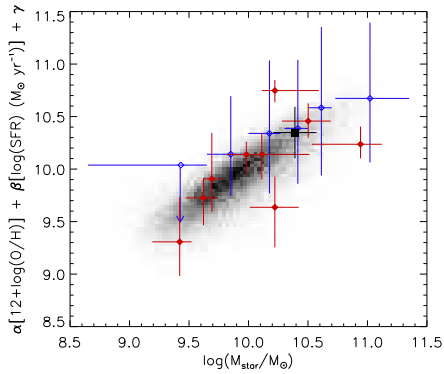}
\caption{FP for field SF galaxies. The stellar mass, in units of solar masses, is presented as a linear combination of 12+log(O/H) and log(SFR). Symbols follow the same code as in Fig 1. Since the \citet{Maiolino08} data do not have SFRs errors, the vertical bars of their data show only the error in metallicity of the linear combination [$\alpha \; \Delta$(12+log(O/H))].}
\end{figure}

We generated least--squares fits of one and two order polynomials combining two of the variables as function of the third one, trying all the possible combinations. As expected, the best one found is a linear combination of the SFR and metallicity yielding the stellar mass (see Fig. 2). The FP for field SF galaxies is then defined by:

{ \small \begin{equation}
{\rm log}({\rm{M_{star}}}/{\rm M_{\odot}}) = \alpha \ [12+\rm log(O/H)] + \  \beta \ [log(SFR) \ (M_{\odot}\rm yr^{-1})] \ + \gamma \ 
\end{equation}} with  $\alpha=1.122 \ (\pm 0.008),\ \beta=0.474 \ (\pm 0.004),\ \gamma=-0.097 \ (\pm0.077)$, and $\sigma=0.16$. The sigma given is that of the vertical axis of Fig. 2.

One of the uses of this FP would be to estimate stellar masses from the metallicity and SFR of emission line galaxies. In analogy to the FP of elliptical galaxies, it is expected that this FP reduces the scatter of the stellar mass found throught the $M-Z$ or mass-SFR relations. To compare the mass (${\rm M_{star}}/{\rm M_{\odot}}$) width of this FP with that of the $M-Z$ and mass--SFR relations, we estimated the 90th and $\sim$68th mass percentile around the fit to each relation, obtaining a mass width for the 90th percentile of $\sim$ 0.57, $\sim$ 0.87, and $\sim$ 0.92 dex for the FP, $M-Z$, and mass--SFR, respectively. Using a $\sim$68th mass percentile, we found a mass range of $\sim$ 0.32, $\sim$ 0.52, and$\sim$ 0.53 dex for the same relations. As observed, the scatter in the mass estimates through the FP is $\sim$ 0.3 and 0.2 dex lower than that of the $M-Z$ relation for the 90th and $\sim$68th mass percentiles, respectively, which means that the mass dispersion using the FP is reduced $\sim$38$\%$ for a 68th percentil.

The largest expected errors of the stellar mass estimates that would be obtained using this FP, without considering SFR and metallicity errors, would be of $\pm 0.28$ dex for a 90th mass percentile, and of $\pm 0.16$ dex for a $\sim$68th mass percentile (1$\sigma$ error). Therefore, as the FP is a well behaved linear relation, and its dispersion in mass is lower than that of the $M-Z$ and mass--SFR relations, it proves to be a useful tool for deriving stellar masses.

In order to identify possible signs of evolution of the FP, we used the data of \citet{Rodrigues08} at z $\sim$0.85, \citet{Erb06} at z $\sim$2.2, and of \citet{Maiolino08} at z $\sim$3.5, as shown in Fig. 1 and 2. The \citet{Erb06} metallicity estimates, using the N2 method and the calibration of \citet{Pettini04}, and the \citet{Maiolino08}, using the calibrations of \citet{Kewley02}, were recalibrated to the R$_{23}$ method using the conversions of \citet{Kewley08}. The \citet{Erb06} data use the \citet{Bruzual03} population synthesis models and a \citet{Chabrier03} IMF. According to \citet{Bruzual03}, the \citet{Chabrier03} and  \citet{Kroupa01} IMF, used by \citet{Erb06} and in this study, respectively, yield practically identical $M/L$ ratios. However, since the stellar masses of \citet{Maiolino08} and \citet{Rodrigues08} are estimated using a \citet{Salpeter55} IMF, we correct them, as indicated by \citet{Maiolino08}, dividing their masses by a factor of 1.17 to make these data consistent with our assumed IMF.

In Fig. 1a the well known evolution of the $M-Z$ relation can be appreciated \citep[e.g.][]{Erb06, Maiolino08}, while in Fig. 1b  an important evolution of the log(SFR) of $\sim$ 1.5 dex is observed for an intermediate mass of 10$^{10.1}$M$_{\odot}$, as well as a flattening in the shape of the mass-SFR relation, already noticed by \citet{Lara10}. The evolution of the $M-Z$ relation, as argued by  \citet{Liang06} is due to a decrease of the metal content in galaxies rather than an increase of their stellar mass. Also, the evolution observed in the mass-SFR relation is due to an increase of SFR in galaxies. Therefore, the metallicity-SFR relation shows evolution in both axes, as shown in Fig. 1c.

We apply Eq. 1 to these high redshift data. As observed in Fig. 2, the Erb et al. (2006) data could show an evolution in the slope. However, the uncertainties of the measured parameters are too large. Moreover, if this change in slope were real, it would be also evident in the data of Maiolino et al. (2008), and it is not. Therefore, we conclude that current high redshift data do not suggest an evident evolution of the FP.

It is noteworthy that the main evolution in the $M-Z$ and mass-SFR relations is driven by the metallicity and SFR, respectively, rather than the stellar mass. Therefore, we would not expect a mass evolution in the mass proyection of the FP, as can be appreciated in Fig. 2. This lack of evolution in our relation could be explained by the metallicity and SFR evolving in opposite direcctions, which means that high redshift galaxies would have lower metallicity values but higher SFRs compared to the local sample. For example, those diferences could be, for an intermediate mass of $\sim$10$^{10.1}$M$_{\odot}$, of 2.12 dex in log(SFR), and of $\sim$0.7 dex in 12+log(O/H). This means that the high SFRs at higher redshifts would be compensated with their lower metallicities when the $\alpha$ and $\beta$ coefficients of Eq. 1. are taken into account.

Stellar mass is the physical fundamental parameter driving the SFR and metallicity of star forming galaxies. Both, SFR and metallicity increase with mass, as shown in the $M-Z$ and mass-SFR relations of Fig. 1. A possible explanation is given by the well known effect of downsizing \citep[e.g.][]{Cowie96,Gavazzi96}, in which the less massive galaxies form their stars later and on longer time scales than more massive systems. This implies lower metallicities and higher specific SFRs for low mass galaxies. Since star formation rate history drives the metal enrichment, downsizing correlates both parameters with mass. Then, their linear combination would relate current star formation rate with its past history, increasing the accuracy of the mass determination by reducing the scatter of the relation.

Usually, the stellar mass of galaxies is estimated through the $z-$band magnitude, the spectral indices $D_n$(4000) and H$\delta_A$, and assuming an IMF, such as the method used by \citet{Kauf03b}. It is also possible to use sophisticated codes such as STARLIGHT \citep{Cid05,Mateus06}, which fits an observed spectrum with a combination of simple stellar populations (SSPs) from the evolutionary synthesis models of \citet{Bruzual03}, computed using a \citet{Chabrier03}  IMF, and $``$Padova 1994" evolutionary tracks \citep{Girardi96}. Moreover, the obtained masses must be corrected for aperture effects based on the differences between the total galaxy magnitude in the $r$ band, and the magnitude inside the fiber, assuming that the mass--to--light ratio does not depend on the radius (see Mateus et al. 2006 for details). Therefore, the use of the FP presented here allows estimating stellar masses in an easier and reliable way compared with the existent methods.


Finally, in an independent and parallel study, \citet{Mannucci10} fit a 2D surface instead of a plane to the same variables, which also reduces the metallicity dispersion. However, one of the most important differences between both studies is that we are using a complete sample in redshift (z$<$0.1) to avoid systematics effects \citep[e.g.][]{Lara10}. Additionally, the plane fitted by us also account for the data at higher redshifts, sustaining the apparent evidence of no-evolution laid down in this section, as explained above, which makes it useful for estimating stellar masses in field galaxies even up to z$\sim$3.5.

\section{Conclusions}

We have demonstrated the existence of a FP for field SF galaxies in the 3D space formed by the orthogonal coordinate axes log(M$_{\rm star}$/ M$_{\odot}$), log(SFR)(M$_{\odot}\rm yr^{-1}$), and 12+log(O/H); three of the fundamental parameters of galaxies.
All those variables have been related previously in pairs as with the $M-Z$, metallicity--SFR, and mass-SFR relations, but this is the first time that the correlation for all of them has been quantified.

The FP presented here allows estimating the stellar mass [log(M$_{\rm star}$/ M$_{\odot}$)] of field galaxies as a linear combination of 12+log(O/H) and log(SFR)(M$_{\odot}\rm yr^{-1}$). The scatter in the mass estimates using the FP (1$\sigma$ error of 0.16) is lower that that obtained through the $M-Z$ and mass-SFR relations.

The FP introduced here would be useful for deriving masses in spectroscopic surveys where the SFR and metallicity are estimated for emission line galaxies, for example, using the H$\alpha$ luminosity to estimate the SFR \citep[e.g.][]{Kennicutt98}, and any of the metallicity methods in the literature, such as the $R_{23}$ \citep{Pagel79} or N2 \citep{Denicolo02}, see \citet{Kewley08} for a review. However, since this study has been carried out using emission line galaxies, this FP will be useful only when both, SFR and metallicity of galaxies can be estimated.

Within the errors, there is no evidence of an evolution of the local FP when applied to high redshift samples. Which means that it could be useful even at high redshifts, where measuring the continuum and absorption lines for fitting models would be more difficult and time consuming.

Then, we propose the use of this FP as an alternative tool to the existing methods to determine the stellar mass of galaxies at low and high redshifts.

\begin{acknowledgements}
This work was supported by the Spanish
\emph{Plan Nacional de Astronom\'{\i}a y Astrof\'{\i}sica} under grant AYA2008-06311-C02-01. We specially thank to C. Bertout, M. Walmsley, F. Moreno-Insertis, and to our last referee for all their help. We thank J. Brinchmann for useful details of MPA-JHU data. M. A. Lara-L{\'o}pez is supported by a CONACyT and SEP Mexican fellowships.

\end{acknowledgements}

\Online

\begin{appendix}

\section{Fundamental plane in a 3D space}

\begin{figure*}[t!]
\centering
\includegraphics[scale=0.90]{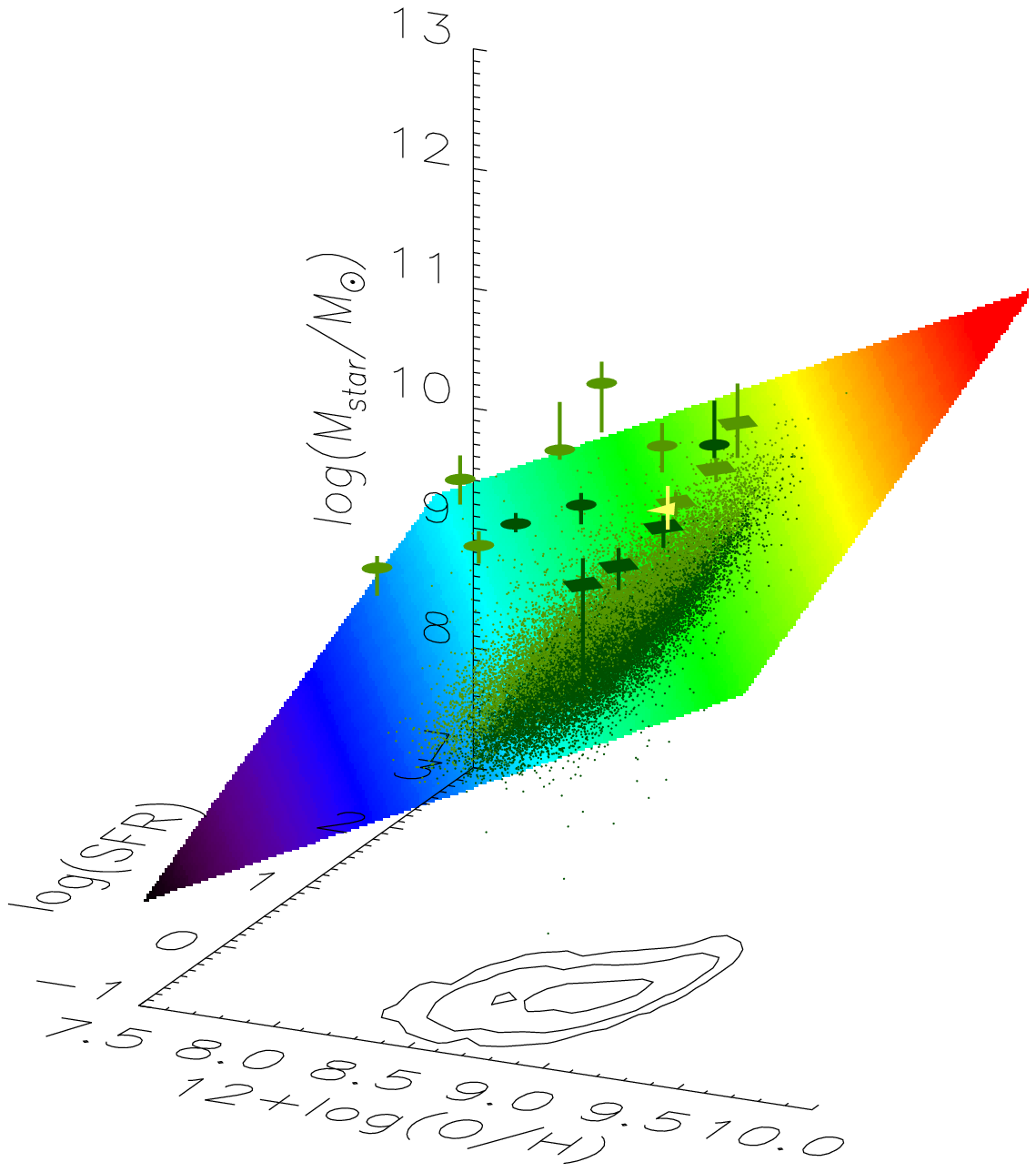}
\includegraphics[scale=0.90]{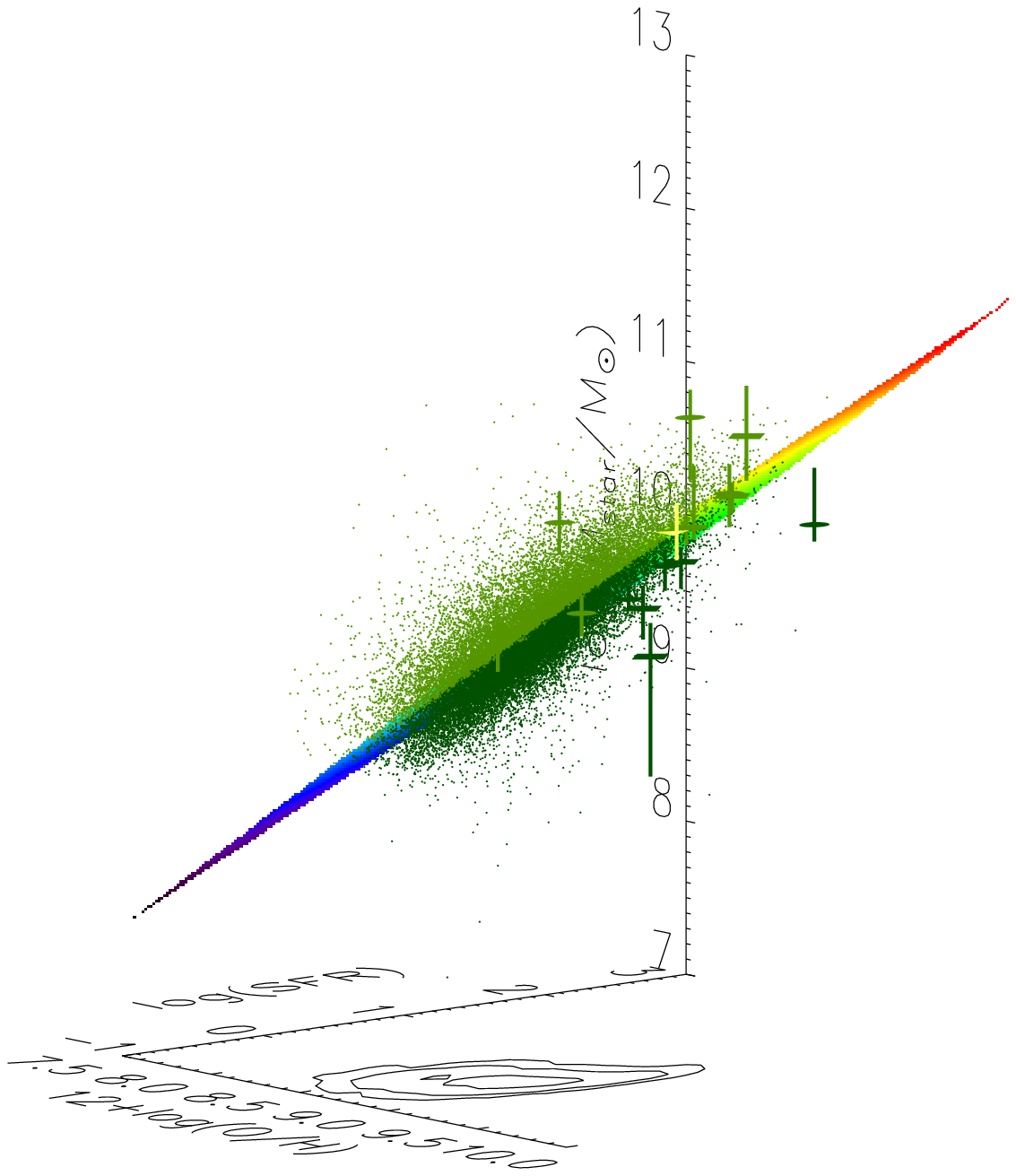}
\caption{Tridimensional representation of the FP according to the fit given in Eq. 1, the color scale of the plane is related with the stellar mass. Light colour symbols are located above the plane, while the dark ones are below it. Dots represent the SDSS local sample described in Sect. 2, diamonds correspond to the Erb et al (2006) data at z$\sim$2.2, and circles symbolise the data at z$\sim$3.5 from Maiolino et al. (2008). The yellow triangle represents the data of Rodrigues et al. (2008) at z$\sim$0.85. The iso-density contours are subject to a projection of the SDSS data cloud on the metallicity-SFR plane, and they are given only as a visual aid.}
\end{figure*}

\end{appendix}

\end{document}